\documentclass[12pt,numbers]{elsarticle}
\pdfoutput=1
\usepackage{amsmath}
\usepackage{bm}

\begin{document}

\begin{frontmatter}

%% Title, authors and addresses

%% use the tnoteref command within \title for footnotes;
%% use the tnotetext command for theassociated footnote;
%% use the fnref command within \author or \address for footnotes;
%% use the fntext command for theassociated footnote;
%% use the corref command within \author for corresponding author footnotes;
%% use the cortext command for theassociated footnote;
%% use the ead command for the email address,
%% and the form \ead[url] for the home page:
%% \title{Title\tnoteref{label1}}
%% \tnotetext[label1]{}
%% \author{Name\corref{cor1}\fnref{label2}}
%% \ead{email address}
%% \ead[url]{home page}
%% \fntext[label2]{}
%% \cortext[cor1]{}
%% \address{Address\fnref{label3}}
%% \fntext[label3]{}

\title{Simulations meet Machine Learning in Structural Biology}

\author{Adri\`{a} P\'{e}rez \fnref{label1}}
\author{Gerard Mart\'{i}nez-Rosell \fnref{label1}}
\author{Gianni De Fabritiis \fnref{label1,label2}}
\ead{gianni.defabritiis@upf.edu}

\address[label1]{Computational Biophysiscs Laboratory (GRIB-IMIM), Universitat Pompeu Fabra, Barcelona Biomedical Research Park (PRBB), Doctor Aiguader 88, 08003 Barcelona, Spain\fnref{label1} }

\address[label2]{Instituci\'{o} Catalana de Recerca i Estudis Avançats (ICREA), Passeig Lluis Companys 23, Barcelona 08010, Spain\fnref{label2} }

%% use optional labels to link authors explicitly to addresses:
%% \author[label1,label2]{}
%% \address[label1]{}
%% \address[label2]{}

\begin{abstract}

%% Text of abstract
 Classical molecular dynamics (MD) simulations will be able to reach sampling in the second timescale within five years, producing petabytes of simulation data at current force field accuracy. Notwithstanding this,  MD will still be in the regime of low-throughput, high-latency predictions with average accuracy. We envisage that machine learning (ML) will be able to solve both the accuracy and time-to-prediction problem by learning predictive models using expensive simulation data. The synergies between classical, quantum simulations and ML methods, such as artificial neural networks, have the potential to drastically reshape the way we make predictions in computational structural biology and drug discovery.
\end{abstract}

\end{frontmatter}
\section*{Highlights}

\begin{itemize}
    \item Classical MD will reach the second timescale and generate petabytes of data within five years.
    \item Accuracy and cost will still restrict simulations to approximate results and high-latency predictions.
    \item Machine learning has the potential to  deliver more accurate and faster predictions by learning  predictive models from expensive simulations. Signs of this paradigm change are already present today in quantum calculations.
\end{itemize}
%% main text
\section*{Introduction}

Molecular dynamics (MD) simulations are one of the predominant techniques to study protein dynamics. MD is often used to capture dynamical processes of proteins across different timescales with atomistic details in order to rationalize biological phenomena.  Despite the potential to become a surrogate model of real protein dynamics, some important issues still remain to be solved, mainly: i) high computational cost and sampling limitations \cite{Freddolino2010} and ii) force field accuracy \cite{Beauchamp2012, Lindorff-Larsen2012, Piana2014}.

Classical MD simulations constitute a balance between accuracy and efficiency. For example, quantum-level phenomena such as enzymatic reactions, polarizability and proton transfers are  neglected in exchange for computational speed. Commonly used  force fields,  based on a parameterization of a closed form potential, are fast to compute, but use approximations that forfeit accuracy. The extent to which these limitations may affect the validity of the results depends on the system and the biological question at hand. Quantum mechanics (QM) calculations can be used to obtain an accurate description of a molecule, but are computationally demanding and very limited in terms of sampling. Ideally, one would like to simulate at quantum level accuracy, which describes the physics and chemistry precisely, but at the sampling regime of current classical simulations.

 The first simulation of protein dynamics dates from 1977 and consisted of a 9.2 ps trajectory of the bovine pancreatic trypsin inhibitor (BPTI) in vacuum \cite{McCammon1977}. In 2010, \cite{Shaw2010} reported a 1 millisecond trajectory of the same protein in explicit solvent, which constitutes a 100 million increase in trajectory length compared to the first simulation. In 30 years, MD simulations have increased sampling capabilities over 8 orders of magnitude, with increasing accuracy in the force fields  \cite{Beauchamp2012, Lindorff-Larsen2012, Piana2014}.  In the last 10 years, MD has evolved from single simulation \cite{Duan1998, Grossfield2008, Dror2009} to high-throughput molecular dynamics studies \cite{Snow2002, Noe2009, Buch2011, Ferruz2015, Pan2017, Stanley2016,Plattner2017}, where hundreds of microseconds  of  simulations  are  computed in  independent trajectories  to obtain  converged statistics. Software and hardware innovations, such as the implementation of MD codes  for  GPUs  \cite{Friedrichs2009, Harvey2009a, Harvey2009, Eastman2017},  distributed  computing projects like Folding@home \cite{Shirts2000}, GPUGRID \cite{Buch2010} and the development of special-purpose supercomputers like ANTON \cite{Shaw2008}, are steadily decreasing the computational cost of molecular simulations.  Additionally, the development of adaptive sampling schemes have introduced more efficient ways to sample conformational space,  decreasing  the  amount  of  simulations  needed  \cite{Singhal2005, Hinrichs2007, Doerr2014}. 

In a recent review we estimated that MD would reach seconds of aggregated sampling using commodity hardware by 2022 \cite{Martinez-Rosell2017} [Fig 1a], generating petabytes of simulation data. For instance, the file size of one second of simulation data of a 60,000-atom system (e.g. a GPCR system) at 0.1 ns per frame is 7.2 Petabytes (reduced to a third using compressed trajectory file formats). This  amount of data constitutes a valuable source of  information, but the knowledge extracted from it is mainly used to rationalize a particular protein system at hand, not to generalize it to other systems. In this review, we envision a paradigm change in the near future where expensive simulations  (QM and MD) are not used to predict but to learn models, so that further predictions can be drawn using ML approaches. By doing so,  the large computational cost required by simulations becomes justifiable, in particular if the results are more accurate by the use of more expensive simulation methods.

\section*{Machine learning applied to structural biology}

ML approaches are not new in simulation analysis. For instance, the common analysis pipeline for MD simulations involves dimensionality reduction \cite{Noe2013, Perez-Hernandez2013, Schwantes2013,Amadei1993, Lange2006, David2014}) and clustering algorithms. 
 
In the last few years,  ML applications have grown exponentially. One of the main factors driving this growth is the broad popularization of a particular type of ML called deep neural networks \cite{LeCun2015, Schmidhuber2015}.  An artificial neural network (NN) is a simple mathematical framework organized in layers, each of them performing a matrix multiplication and a non-linear function of the input variables $\boldsymbol{x}$. The output of a single neuron $\phi$ in each layer  is given by $\phi= f(\boldsymbol{w}^t  \boldsymbol{x} +b)$, where $\boldsymbol{w}$ are learnable weights,  \textit{b} is a bias and \textit{f} is some nonlinear function.  NNs can have several to hundred of nested layers and in such cases is called ''deep''. Given enough parameters, a NN is capable of interpolating any continuous function \cite{Hornik1991, Andoni2014}.
 
The application of NN models  in computational biology is steadily increasing \cite{Angermueller2016}. For instance, the Merck molecular  activity  challenge  demonstrated  the  potential of deep neural network models in drug discovery   \cite{Dahl2014}. DeepTox \cite{Mayr2015} is a deep learning-based model for toxicity prediction of compounds, winning the Tox21 toxicology prediction challenge in 2014 by a large margin. Variational autoencoders \cite{Kingma2013}, a generative flavor of deep NNs, were recently applied to convert  discrete  representations  of  molecules  to  and  from  a  multidimensional continuous representation \cite{Gomez-Bombarelli2016}, allowing for efficient search and optimization through  open-ended  spaces  of  chemical  compounds. Additionally, autoencoders have also been used for dimensionality reduction in MD \cite{Wehmeyer2017, Doerr2017, Hernandez2017}. VAMPnets \cite{Mardt2017}  fit a Markov state model from the system specific molecular simulation data.  NNs have also been used to reproduce the free-energy surface of molecules \cite{Schneider2017}. Deep convolutional neural networks (CNN) \cite{Krizhevsky2012} have become increasingly popular due  to  its   performance  in  machine  vision,  a property that has been  exploited by us and others to apply it on structural biology by treating proteins as 3D images. CNNs have been used for ligand binding site detection \cite{Jimenez2017}, ligand pose prediction \cite{Ragoza2017}, ligand active/inactive classification \cite{Wallach2015}, ligand binding affinity prediction \cite{Jimenez2018} and protein design \cite{Torng2017}. Also, the DeepChem software \cite{deepchem} and the MoleculeNet challenge \cite{Wu2017} provide  multiple  featurization  algorithms  and  access to relevant QSAR prediction datasets. 

\begin{figure*}[!t]%figure1
\includegraphics[width=\textwidth]{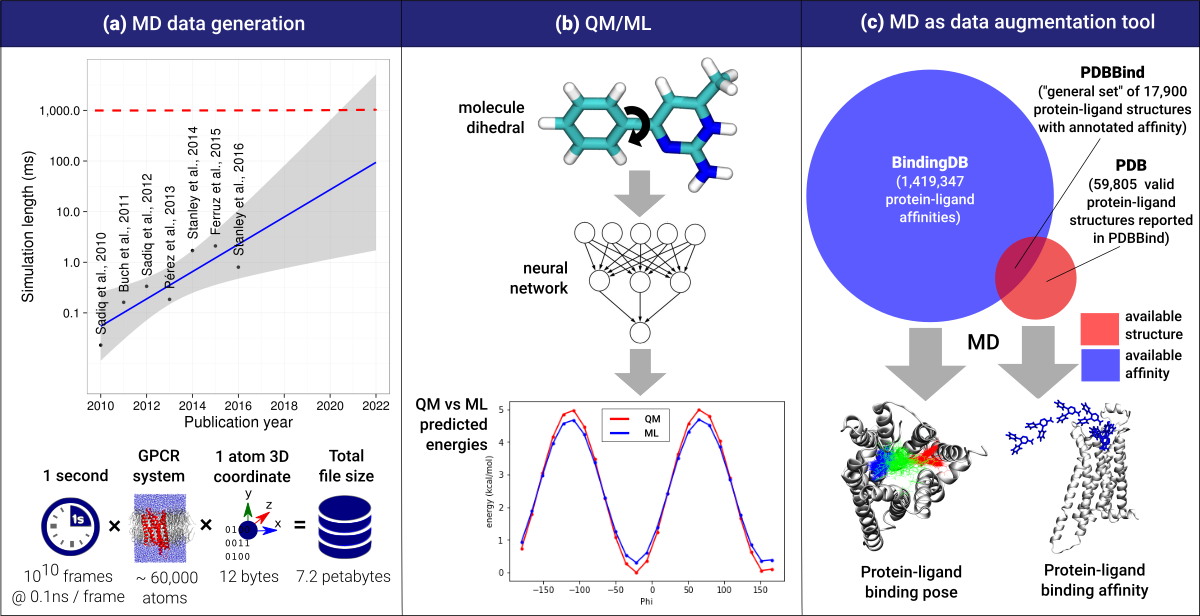}
\caption{Overview of a combined simulation and machine learning approach. \textbf{a}.  MD data generation is expected to reach the second aggregated timescale by 2022 and an output files size of several petabytes by 2022 based on a trend of  maximum aggregated time per paper per year using the ACEMD software. Chart adapted from \cite{Martinez-Rosell2017}. Referenced publications correspond to \cite{Buch2011, Ferruz2015, Stanley2016, Perez-Hernandez2013, Sadiq2010, Sadiq2012,  Stanley2014}. \textbf{b}. A first example of ML replacing QM to predict dihedral energies given a neural network trained with QM simulations. \textbf{c}. An example of data augmentation by MD: augment protein-ligand binding poses for a set of protein-ligand pairs with unknown binding mode; augment binding affinity data for a set of resolved protein-ligand complex structures of unknown affinities.}\label{fig:01}
\end{figure*}

\section*{Case 1: ML models to predict quantum forces using QM simulation data}

One important application case of ML is apparent in quantum mechanics. QM simulations are notoriously computationally expensive and, depending on the level of theory,   scales  poorly with the number of atoms of the system \cite{Carleo2017}. It is therefore not surprising that there have been efforts to interpolate the QM many-body potential with NNs to obtain a predictive model of QM forces. 

Many studies on approximating QM with NNs were performed previously to the recent NN resurgence. In particular, Behler et al. \cite{Behler2007, Behler2011, Behler2015} contributed significantly to the field for small molecules and on ways to fit quantum observables, e.g.  infrared spectra  \cite{Gastegger2017}. The initial  effort went to provide usable symmetry functions that could guarantee basic physical principles like translational, rotational \cite{Boomsma2017} and invariance on atom reordering to the learned potentials. Transferability, however, was limited until recently\cite{Smith2017}. 

In \cite{Smith2017, Yao2017, Zhang2017, Faber2017}, NNs are trained from QM simulation data to generate the potential energy surface and forces for small molecules,  generalizing to unseen molecules, including some preliminary tests on proteins. In the same way as MD force fields do,   forces are true derivatives of the interpolated potential energy surface using the gradients of the NN, and can be used to run dynamics. This guarantees that the forces produced by the NNs yields a conservative field \cite{Chmiela2017}. The QM energy potential is therefore learned  with the accuracy of first-principle based methods, using generated datasets for many molecules.  The computational cost of generating the datasets is of course very large, but once trained, the NN inference cost is many orders of magnitude faster than the QM computational model and comparable in costs to standard classical MD [Fig 1b].

\section*{Case 2: ML models to predict binding affinities using MD simulation data}

MD software for GPU  has made simulations of full protein-ligand binding processes possible, allowing the prediction of thermodynamic and kinetic properties \cite{Buch2011, Ferruz2015}. At the moment, a trade-off between accuracy and  sampling restricts the applicability of MD compared to other commonly used methods used in drug design, like docking, less accurate but significantly faster. Even if the sampling problem is solved via brute force, MD does not currently guarantee that the results are correct because of the approximations of the force fields. The last point can be mitigated by the use of QM/ML force fields in the future.

Recently, we explored the use of machine vision NN models for binding affinity prediction. In  \textit{K}\textsubscript{DEEP} \cite{Jimenez2018}, a ML model is used to predict binding affinities, which consists of a CNN trained on the PDBBind database \cite{Liu2017}. The method shows an overall good mean correlation (0.82) when tested against the PDBBind's core set. This set contains several targets clustered by sequence similarity, in order to define a representative, non-redundant subset of proteins. For few of these protein clusters, however, the correlation disappears or even becomes negative. This fact might be explained by a lack of training data for specific protein pockets, which ultimately leads to a poor generalization in these cases. 
 \textit{K}\textsubscript{DEEP} is structure-based, i.e. it requires labelled data in the form of the structure of the protein-ligand complex and their affinity. One way to address this issue would be to extend the available training datasets by obtaining new affinity data or structures, either experimentally or computationally (Fig. \ref{fig:01}c). Experiments are of course a possibility, viable for pharmaceutical companies and some academic groups. Here we prefer to look at the computational options which can be more automated in active learning methods and are subjected to be exponentially cheaper in the future. 
 
 A potential synergy between MD and ML would improve the accuracy of predictive NN models, delivering predictions several orders of magnitude faster than simulations. This level of performance is needed for large prediction studies in drug design, where thousands of molecules need to be evaluated, such as in virtual screening. As for training \textit{K}\textsubscript{DEEP}, the two most popular binding affinity databases are PDBBind \cite{Liu2017} and Binding MOAD \cite{Ahmed2015}. PDBind's latest release (v2017) screened the 124,962 structures in the PDB database \cite{Berman2000} (as in Jan 1st, 2017) and identified 59,805 valid molecular complex structures into four main categories: protein-small ligand, nucleic acid-small ligand, protein-nucleic acid and protein-protein complexes.  From this set of structures, they defined the general set,  providing binding affinity data (KD/KI and IC50) for a total of 17,900 biomolecular complexes in the PDB database, including protein-ligand (14,761), nucleic acid-ligand (121), protein-nucleic acid (837), and protein-protein complexes (2,181). The other dataset, Binding MOAD, contains binding information for 9142 structures, being 6862 of them overlapped with PDBBind. This makes a total of 20,065 co-crystal structures with binding data, out of the 59,805 complex structures detected in the PDB. A naive example of synergy could be to increase the available affinity data for the remaining 39,740 structures by simulations. This, however would be very expensive and arguably impractical in a prediction study. Yet in the context of generating a database for training NN models, it only needs to be performed once, and possibly at very high accuracy using QM/ML-based force fields to obtain very accurate data. The resulting NN will be used for predictions. Another possible example comes from the BindingDB dataset \cite{Gilson2016} which contains about 1,419,347 binding data for 7,000 proteins and over 635,301 drug-like molecules, but for most of the protein-ligand pairs there is  no co-crystal structural information. To fill up this gap, simulations could be used to predict  ligand binding poses. As a rough estimation, if approximately 10$\mu$s are needed to obtain the ligand binding pose for a pair of protein-ligand using adaptive sampling methods\cite{Doerr2014}, with 1s of aggregate time one could generate 100,000 new predicted protein-ligand structures over the course of one year\cite{Martinez-Rosell2017}. This augmented dataset build at high computational  and time cost is then used for learning fast predictive models, e.g. \textit{K}\textsubscript{DEEP}.

\section*{Discussion}

In this article we illustrate how generated data produced by simulations might be used  to develop new and better predictive ML models. Generation of datasets is not hampered by fast return times, which means that better simulation methods can be used, while ML is used to obtain fast predictions. One existing example of this approach are  QM simulations of biomolecules, used to generate data for learning a NN QM potential, a paradigm that could improve on classical force fields in the near future. A further possible example, build upon the experience obtained in \textit{K}\textsubscript{DEEP},  is where  simulations are used as a data augmentation tool, delegating the binding affinity prediction to ML-based methods.

\section*{Acknowledgements}
  
The authors thank Acellera Ltd. for funding. G.D.F. acknowledges support from MINECO (BIO2017-82628-P) and FEDER, as well as 'Unidad de Excelencia Mar\'{i}a de Maeztu', funded by MINECO (MDM-2014-0370). The authors thank the European Union’s Horizon 2020 research and innovation programme under grant agreement No 675451 (CompBioMed project).
\section*{References}

\appendix
%% The Appendices part is started with the command \appendix;
%% appendix sections are then done as normal sections
%% \appendix

%% \section{}
%% \label{}

%% If you have bibdatabase file and want bibtex to generate the
%% bibitems, please use
%%
%%  \bibliographystyle{elsarticle-harv} 
%%  \bibliography{<your bibdatabase>}

%% else use the following coding to input the bibitems directly in the
%% TeX file.

\bibliographystyle{ieeetr}
\bibliography{COSBpaper}{}

\end{document}